\documentclass[9pt,letterpaper,twocolumn]{article} 
\usepackage[T1]{fontenc} 
\usepackage[latin1]{inputenc}

\usepackage[draft]{hyperref} 
\usepackage{graphicx} 
\usepackage{srcltx}
\usepackage{setspace}
\usepackage{amsmath}

\begin{document}

\twocolumn[

\title{Theoretical study of Acousto-optical coherence tomography
using random phase jumps on US and light}

\author{M. Lesaffre$^{1,2}$, S. Farahi$^1$,
A.C. Boccara$^1$, F. Ramaz$^1$ and M. Gross$^{2,3,*}$\\ \\
 $^1$Institut Langevin, ESPCI ParisTech, CNRS UMR 7587,\\ 10 rue Vauquelin
F-75231 Paris Cedex
05.\\
$^2$Laboratoire Kastler-Brossel de l'ENS, UMR 8552 (ENS, CNRS, UMPC), \\24  rue Lhomond
F-75231 Paris Cedex
05\\
$^4$Laboratoire Charles Coulomb  UMR 5221 CNRS-UM2 Universit\'{e} Montpellier \\II place
Eug\`{e}ne Bataillon 34095 Montpellier \\
$^*$Corresponding author: gross@lkb.ens.fr}

\maketitle

\begin{abstract}
Acousto-Optical Coherence Tomography (AOCT)    is  variant of Acousto Optic Imaging
(called also ultrasonic modulation imaging) that makes possible to get $z$ resolution
with acoustic and optic Continuous Wave (CW) beams. We describe here theoretically the
AOCT effect, and we show that the Acousto Optic ''tagged photons'' remains coherent if
they are generated within a specific $z$ region of the sample. We quantify  the $z$
selectivity for both the ''tagged photon'' field, and for the M. Lesaffre et al.
photorefractive signal.
\end{abstract}


\bigskip


\bigskip

]

\section{Introduction}

Acousto-optic imaging (AOI) \cite{kempe1997acousto,leveque2001three,atlan2005pulsed} is a
technique that couples ultrasounds and light in order to reveal the local optical
contrast of absorbing and/or scattering objects embedded within thick and highly
scattering media, like human breast tissues.
%

First experiments used fast  single detectors  to record the  modulation of the optical
signal at the US frequency
\cite{wang1995continuous,leutz1995ultrasonic,wang1997ultrasound,kempe1997acousto,yao2000theoretical}.
But, since the phase of the modulation is different for each grain of speckle, the
detector can only process one grain of speckle. To increase the optical etendue of
detection, Leveque et al. \cite{leveque1999utp} have developed a camera detection
technique that processes many speckles in parallel. This technique has been pulled to the
photon shot noise limit by Gross et al. \cite{gross2003snd} using a holographic
heterodyne technique \cite{leclerc2000nhh} able to detect photons with optimal
sensitivity \cite{gross2007dhu,verpillat2010digital}.
Since the US attenuation is low in tissues, the tagged photons are generated along the US
propagation $z$ axis with a nearly constant rate. This means that in a continuous regime
of the US, the AO techniques give nearly no information on the location of the embedded
objects along the $z$ axis. To get such $z$ information, Wang et al. \cite{wang1998fsu},
have developed  a US frequency chirp technique with a single detector, which has been
extended to camera detection \cite{yao2000frequency,forget2003high}. Unfortunately, these
chirp techniques cannot be used in living tissues, because the phase of light
decorrelates very fast in them, since  half frequency linewidth of light that travels
through 4 cm of living breast tissue is about 1.5 kHz \cite{gross2005hdm}.  This phase
decorrelation drastically lowers the detection efficiency,  since the detection bandwidth
is approximately equal to the camera frame rate. The bandwidth is then much narrower than
the width of the scattered photon frequency spectrum, and most of the tagged photons are
undetected. It is still possible to increase the detection bandwidth by using a faster
camera, but in such systems this generally means that a smaller number of pixels should
be used, and the optical etendue of detection decreases accordingly.
To perform selective detection of the tagged photons with high optical etendue, narrow
band incoherent detection techniques have been proposed. For example, Li et al. select
the tagged photon by spectral holeburning \cite{li2008pulsed,li2008detection}, while
Rousseau uses a confocal Fabry-Perot interferometer \cite{rousseau2009ultrasound}. This
last experiment \cite{rousseau2009ultrasound} benefits from  a powerful long pulse laser,
whose duration (0.5 ms) matches the 1.5 kHz signal bandwidth.
Another way  to get a detection bandwidth comparable  with the  signal bandwidth while
keeping a large optical etendue, detection schemes involving photorefractive (PR)
crystals have been  proposed. Murray et al. \cite{murray2004detection,sui2005imaging} use
a PR crystal sensitive at $ 532$ nm to select the untagged photons, which are detected by
a single avalanche photodiode . In this case, the weight of the tagged photon signal is
measured indirectly by using the conservation law of the total number of photons (tagged
+ untagged) \cite{gross2005theoretical}. Ramaz et al. selectively detect either the
tagged or the untagged photons \cite{ramaz2004photorefractive}. The Ramaz technique is
also able to measure in situ the photorefractive writing time ($\tau_{PR}$),
which characterizes the detection frequency bandwidth \cite{lesaffre2007smp}. 

In order to get information on the location  of the object along the $z$ axis, acoustic
pulses can be used. The method has been extensively used both with single detectors
\cite{lev2003pulsed,lev2005ultrasound}, cameras \cite{atlan2005pulsed},  PR crystals
without
\cite{murray2004detection,sui2005imaging,farahi2010photorefractive,bossy2005fusion}, or
with long pulse laser \cite{rousseau2008ultrasound}. Nevertheless, reaching a millimetric
resolution with US pulses requires a typical duty cycle of $1\%$, corresponding to the
exploration length within the sample ($\sim 10$ cm) and the desired resolution ($\sim 1$
mm). This is problematic regarding the very small quantity of light that emerges from a
clinical sample, since weak duty cycle yields low signal and poor Signal-to-Noise Ratio
(SNR). When US pulses are used with photodiode detection, the SNR becomes lower, since
fast photodetectors mean larger electronic noise. In a recent publication, Lesaffre et
al. \cite{lesaffre2009acousto} overcome the duty cycle problem, and  get  $z$ resolution
with CW light and ultrasound by applying a random phase modulation on both the optical
illumination and US beam. This so called Acousto-Optical Coherence Tomography (AOCT)
technique is then demonstrated with photorefractive detection of the tagged photons.

Whatever the method used to obtain an axial resolution, the acousto-optic signal is
sensitive to the quantity of photons tagged by the ultrasound : as shown in many previous
studies, a strong absorber ("zero" transmission") within the US field will induce an
important drop on the signal
\cite{lesaffre2009acousto,kempe1997acousto,gross2007dhu,li2008detection,rousseau2009ultrasound,gross2005theoretical}.
It has been shown more recently that a small "quasi-transparent" inclusion having a
scattering coefficient ($\mu'_s=10$ cm$^{-1}$) different from the host matrix ($\mu'_s=7$
cm$^{-1}$) can give a contrast in the acousto-optic signal \cite{lai2009quantitative}. In
both cases, and to our knowledge, no quantitative measurements of this contrast have been
performed as a function of the absorption coefficient nor the transport mean free path
length $l^*$.

In the present paper, we will describe the AOCT effect theoretically. We show that the
tagged photons remain coherent if they are generated within a specific $z$ region of the
sample.  We will quantify  the $z$ selectivity for both the tagged photon field, and for
the tagged photon photorefractive  signal as detected by Lesaffre et al.
\cite{lesaffre2009acousto}. The theoretical results we get here will be compared with
experiment in another publication.

\section{Theory of the Acousto Optic Coherent Tomography (AOCT).}

The theoretical description of the Acousto  Optic Coherent Tomography  cannot be simply
extrapolated from the theory made previously \cite{gross2005theoretical} to describe  the
photorefractive detection of the UltraSound Modulated photons (USM). Since we make
tomography,  we cannot consider that the USM photons are globally generated by the
modulation of the length of a travel path. We must  make a finer analysis by describing
how the USM photons are locally generated within each specific region of the sample.

\subsection{The generation of the "tagged photons"}\label{section_1}

\begin{figure}[]
\centering{}
\includegraphics[width=6cm]{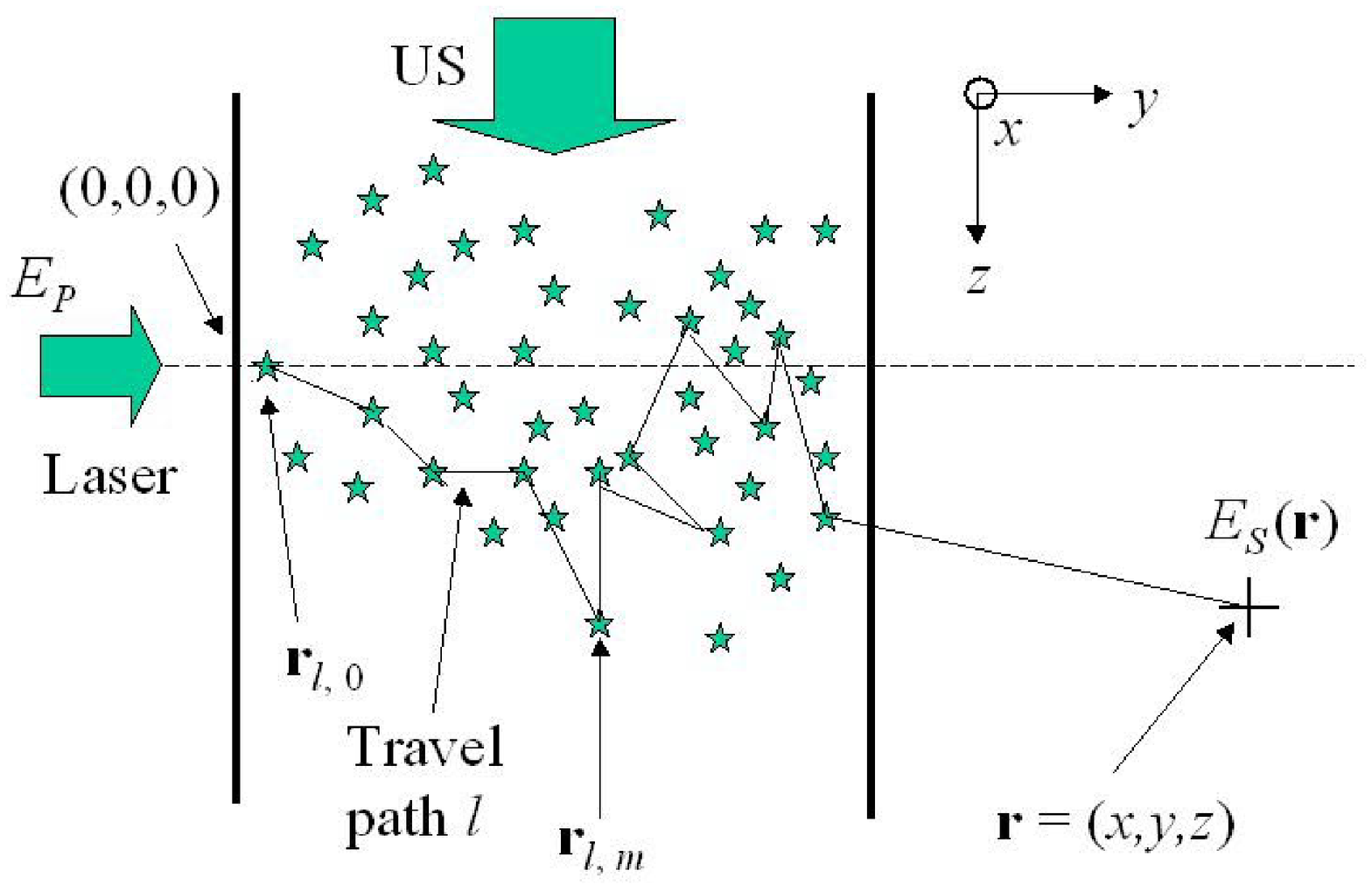}
\caption{Light scattering along the travel path $l$, which involves scattering
event located in $\textbf{r}_{l,m}$, where $m$ is the scattering event index.} \label{fig_principle}
\end{figure}

Let us call  $E_P$ and $E_S$ the fields coming into and out  of the sample.  Consider the
point ($x,y$) located after the sample output interface. $E_S$ is a quasi monochromatic
wave at the frequency $\omega_0$ of the incoming laser. Let's introduce the complex
field amplitude ${\underline E}_P$ and ${\underline E}_S$ defined as:
\begin{equation} \label{Eq_1}
E_{P}(t)=\Re\left\{ {\underline{E}}_{P}e^{j\omega_{0}t}\right\}
\end{equation}
\begin{equation}\label{Eq_2}
E_{S}(t)=\textbf{\ensuremath{\Re}}\left\{ {\underline{E}}_{S}e^{j\omega_{0}t}\right\}
\end{equation}
%
%
%
where $\textbf{\ensuremath{\Re}}$  is the real part operator.  $E_S$ results from the sum
(or the interference) of the field components $E_{S,l}$ scattered through the sample
along many travel paths $l$ from input plane ($z = 0$) to the detector.  Moreover, as
illustrated by Fig.\ref{fig_principle},  each travel path $l$ can be decomposed in a
succession of scattering events ($l,m$) located in $\textbf{r}_{l,m}$ where $m$ is the
scattering events index.
%
%
%
\begin{eqnarray}\label{Eq_4}
E_{S}\left(t\right)&=&\sum_{l} E_{S,l}\left(t\right)\\
\nonumber
&=&\textbf{\ensuremath{\Re}}\left\{ \sum_{l}~a~{\underline{E}}_{P}~
e^{j\omega_{0}\left(t-{s_{l}}/{c}\right)}\right\}
\end{eqnarray}
%
%
%
%
where $l$ is the travel path index, and $s_l$ the corresponding effective travel path
length. The length $s_l$  is the product of the travel path length by the medium
refractive index $n$. To simplify the discussion we  consider that the field amplitude
$a$ is the same for all the travel paths. All travel paths have the same weight
$a{\underline{E}}_{P}$, but different field phases: $e^{-j~\omega_{0}s_{l}/c} \equiv
e^{-j~2\pi s_{l}/\lambda}$. Since the travel path lengthes are large with respect to the
optical wavelength $\lambda$, the factor $e^{-j{2\pi s_{l}}/{\lambda}}$ is random.
Summing over the travel paths,  one gets a speckle outgoing field.

\subsubsection{The ultrasonic field of pressure}

Let us now apply a CW (Continuous Wave) ultrasonic (US) wave to the system by using an
ultrasonic piezoelectric (PZT) device. The PZT transducer excitation voltage is:
\begin{equation}\label{Eq_5}
U_{PZT}(t)=\textbf{\ensuremath{\Re}}\left\{ {\underline{U}}_{PZT}\;e^{j\omega_{US}t}\right\}
 \end{equation}
where $\underline{U}_{PZT}$ is the complex amplitude of $U_{PZT}$. Like in experiments,
we consider here linear conditions where the acoustic pressure $P_{US}$ is proportional
to the excitation voltage. By this way, we get in any point $\textbf{r}$ of the sample:
\begin{equation}\label{Eq_6}
P_{US}(\textbf{r},t)=A(\textbf{r})\;{U}_{PZT}\left(t-{z}/{c_{US}}\right)
\end{equation}
where $c_{US}$ is the sound velocity in the sample, and ${z}/{c_{US}}$ the time delay
from the US emission point (the PZT)  to the zone of coordinate $z$ that is considered.
Let us introduce the US pressure complex amplitude ${\underline{P}}_{US}$:
%
\begin{eqnarray}\label{Eq_7}
P_{US}(\textbf{r},t)&=&\Re~\left\{ {\underline{P}}_{US}(\textbf{r})~e^{j\omega_{US}t}\right\}
\end{eqnarray}
with
\begin{eqnarray}\label{Eq_7_a}
{\underline{P}}_{US}(\textbf{r})&=&A(\textbf{r})~{\underline{U}}_{PZT}~e^{{-j\omega_{US}z}/{c_{US}}}
\end{eqnarray}
The pressure  $\underline{P}_{US}$  is periodic with respect to the US propagation axis,
the period being $\lambda_{US}=2\pi{c_{US}}/{\omega_{US}}$.

\subsubsection{The acousto optic modulation}

Because of the US beam  the scatterers  vibrate. Moreover,  the sample refractive
index is modulated. These two effects yield a modulation of the length $s_l$ of the
travel paths of the photons that are scattered by the medium (where $l$ is the travel
path index) at the US frequency  $\omega_{US}$:
\begin{equation}\label{Eq_8}
s_{l}(t)=s_{l,0}-\Re\left\{ \underline{\delta s}_{l}\;e^{j\omega_{US}~t}\right\}
 \end{equation}
where $\underline{\delta s}_{l}$ is the  complex amplitude of the modulation  of the
travel path $l$. We get from Eq.\ref{Eq_4}:
\begin{eqnarray}\label{eq:8a}
 E_{S}\left(t\right)&=&\textbf{\ensuremath{\Re}}
\sum_{l}~a~{\underline{E}}_{P}~e^{j~\omega_{0}\left(t-{s_{l,0}}/{c}\right)}
~\\
\nonumber
&& \times
\exp\left[j~\frac{\omega_{0}}{c}\Re\left\{ \underline{\delta s}_{l}\; e^{j\omega_{US}~t}\right\} \right]
\end{eqnarray}
Let us introduce the complex amplitude   $\underline{\delta s}_{l,m}$ of  the
$\textrm{m}^{th}$ scatterer  contribution to the travel path modulation, whose modulus
and phase are $\beta_{l,m}$ and $\phi_{l,m}$ respectively.
%
\begin{eqnarray}\label{xxx}
\underline{\delta s}_{l}=\sum_{m}\underline{\delta s}_{l,m}=\sum_{m}
  \beta_{l,m}e^{j\phi_{l,m}}
\end{eqnarray}
The sample outgoing field  $E_{S}(t)$ is then modulated by the  US at frequency
$\omega_{US}$.
\begin{eqnarray}\label{Eq_10}
E_{S}(t)&=& a\;\Re \;\sum_{l}\;{\underline{E}}_{P}\;e^{j\;\omega_{0}
\left(t-{s_{l,0}}/{c}\right)}~\\
\nonumber
 && \times \exp\left[j~\frac{\omega_{0}}{c}\sum_{m} \Re\left[~\underline{\delta
s}_{l,m}e^{j\omega_{US}t}\right]\right]
\end{eqnarray}

\subsubsection{The tagging of the scattered photons}
%

In typical experiments, the  vibration amplitude is much lower than the optical
wavelength $\lambda=2\pi{c}/{\omega_{0}}$: for example,   the vibration amplitude is  60
nm for 1 MPa  acoustic pressure at $\omega_{US}=2$ MHz.  We can then make the hypothesis
of a weak acousto optic modulation:
%
\begin{equation}
\frac{\omega_{0}}{c}\sum_{m}\Re\left\{ \underline{\delta s}_{l,m}e^{j\omega_{US}t}\right\} ~\ll~1
\label{Eq_11}
\end{equation}
We get in Eq.\ref{Eq_10}:
\begin{eqnarray}\label{Eq_12}
 && \exp\left[j~\frac{\omega_{0}}{c}\sum_{m}\Re\left\{ \underline{\delta
s}_{l,m}e^{j\omega_{US}t}\right\}
\right] \simeq \\
\nonumber
&&~~~~~~~~ 1+j~\frac{\omega_{0}}{c}\sum_{m}\Re\left\{ \underline{\delta s}_{l,m}e^{j\omega_{US}t}\right\}
\end{eqnarray}
The field $E_{S}(t)$ diffused by the sample becomes:
%
%
%
\begin{eqnarray} \label{Eq_10a}
\nonumber
E_{S}(t)  =  \Re \left\{\left[ a\;\sum_{l}\;{\underline{E}}_{P}\;e^{j\omega_{0}\left(t-{s_{l,0}}/
{c}\right)}\right] \right.\\
\left. \times \left[1+ j~\frac{\omega_{0}}{c}\sum_{m}\Re\left\{ \underline{\delta s}_{l,m} e^{j\omega_{US}t}\right\}\right]\right\}
 \end{eqnarray}
The  field $E_{S}(t)$ diffused by the sample is the sum of a main component
$E_{S,\omega_{0}}(t)$, whose frequency is $\omega_{0}$, with the two sideband components
$E_{S,\omega_{\pm1}}(t)$, whose frequencies are $\omega_{\pm1}=\omega_{O}\pm\omega_{US}$.
\begin{eqnarray} \label{Eq_14}
E_{S}(t)&=&E_{S,\omega_{0}}(t)+E_{S,\omega_{1}}(t)+E_{S,\omega_{-1}}(t)
\end{eqnarray}
Let us introduce ${\underline{E}}_{S,\omega_{0}}$ and
${\underline{E}}_{S,\omega_{\pm1}}$, which are slow varying with time.
\begin{eqnarray} \label{Eq_14}
E_{S,\omega_{0}}(t)& \equiv&\Re~\left\{ {\underline{E}}_{S,\omega_{0}}\exp(j\omega_{0}t)\right\}\\
\nonumber   E_{S,\omega_{\pm1}}(t)&\equiv&\Re~\left\{ {\underline{E}}_{S,\omega_{\pm1}}\exp(j\omega_{\pm1}t)\right\}
\end{eqnarray}
%
We get from Eq.\ref{Eq_10a}:
\begin{eqnarray}\label{Eq_14xx}
E_{S,\omega_{0}}(t)  =  \Re\left\{ \;
a\;\sum_{l}\;{\underline{E}}_{P}\;e^{j\;\omega_{0}\left(t-{s_{l,0}}/{c}\right)}\right\}
\label{Eq_11b}\end{eqnarray}
\begin{eqnarray}\label{Eq_20}
&&E_{S,\omega_{1}}(t)+E_{S,\omega_{-1}}(t) =  \Re~\left\{ a{\underline{E}}_{P}
e^{j\omega_{0}t}
\vphantom{\sum_{l,m}\left[j\frac{2\pi\beta_{l,m}}{\lambda}~e^{-j{2\pi s_{l,0}}/{\lambda}}
  \left[e^{j\phi_{l,m}}\; e^{~j\omega_{US}t}+c.c.\right]~\right] }
\right.\\
\nonumber &&   \left. \times \sum_{l,m}\left[j\frac{2\pi\beta_{l,m}}{\lambda}~e^{-j{2\pi s_{l,0}}/{\lambda}}
  \left[e^{j\phi_{l,m}}\; e^{~j\omega_{US}t}+c.c.\right]~\right]\right\}
\end{eqnarray}
where $c.c.$ means the complex conjugate. We thus have for 
${\underline{E}}_{S,\omega_{\pm1}}$:
%
\begin{eqnarray}\label{Eq_22}
{\underline{E}}_{S,\omega_{\pm1}}(t)&=&a{\underline{E}}_{P} \sum_{l}
 \\
\nonumber && \left[j~e^{-j{2\pi
s_{l,0}}/{\lambda}} \times\sum_{m}\left[\frac{2\pi\beta_{l,m}} {\lambda}e^{\pm j\phi_{l,m}}\right]\right]
\end{eqnarray}
Here, the main component $E_{S,\omega_{0}}$ does not depend on  the travel path
modulation (Eq.\ref{Eq_14xx}), while the modulated components $E_{S,\omega_{\pm 1}}$ do.
Moreover, whatever the modulation mechanism is: displacement of the scatterers or
modulation of the refractive index, $\beta_{l,m}$ is directly related to the acoustic
pressure $P_{US}\left(\textbf{r}_{l,m}\right)$ at the scatterer location
$\textbf{r}_{l,m}$.

Note that the phases $\phi_{l,m} $ and $\phi_{l,m'} $ of two  scattering events $m$ and
$m'$ of the same path $l$ are partially correlated according to the position of the
associated diffusers $\textbf{r}_{l, m} $ and $\textbf{r}_{l,m'} $, and according to the
physical effect at the origin of the modulation.

For the displacement of the scatterers, the  phases $\phi_{l,m} $ is related to the
projection $q_z$ of the scattering wave vector $\textbf{q}_{l,m}$ along the US
propagation direction (i.e $z$) with $\textbf{q}_{l,m}=\textbf{k}'_{l,m}-
\textbf{k}_{l,m}$ (where $\textbf{k}_{l,m}$ and $ \textbf{k}'_{l,m}$ are the wave vectors
of the photon before and after the scattering event $l,m$). The phases $\phi_{l,m} $ and
$\phi_{l,m'} $ are not correlated, since $q_z$ may change of sign from one scattering
event ($l,m$) to the next ($l,m+1$) within the same path $l$.

For the modulation of the refractive index,  $\phi_{l,m} $ is mainly related to the US
phase. In a typical experiment the scattering length $l_s$ is about $0.1$ mm, while the
US wavelength $\lambda_{US}$ is about 1 mm (0.75 mm for $\omega_{US}=2$ MHz). This means
that  $\phi_{l,m} $ and $\phi_{l,m'} $ are correlated, if the scattering events ($l,m$)
and ($l,m'$) are close together ($|m-m'|<$a few units), and uncorrelated if not.

This partial coherence allows us to use the acousto-optical modulation in scattering
media. However, all the scatterers $\textbf {r}_{l, m}$ in the acoustic column contribute
to the tagged photons  field ${\underline{E}}_{S,\omega_{\pm1}}(t)$. Thus on the acoustic
column, the information is not localized. So it is  necessary to use a complementary
technique in order to obtain an axial $z$ resolution.

\subsection{The axial resolution along $z$}

\label{section_z_selction}
%

To obtain an axial resolution along $z$, Lesaffre et al. \cite{lesaffre2009acousto} have
used Acousto Optic Cohérent Tomography (AOCT). This technique is based on the control of
the acoustic and optical coherence lengths using a random phase modulation on the
acoustic and
optical arms.  

\subsubsection{The AOCT random modulation   of the optical and acoustical field phases.}

\begin{figure}[h]
\centering{}\includegraphics[width=8cm]{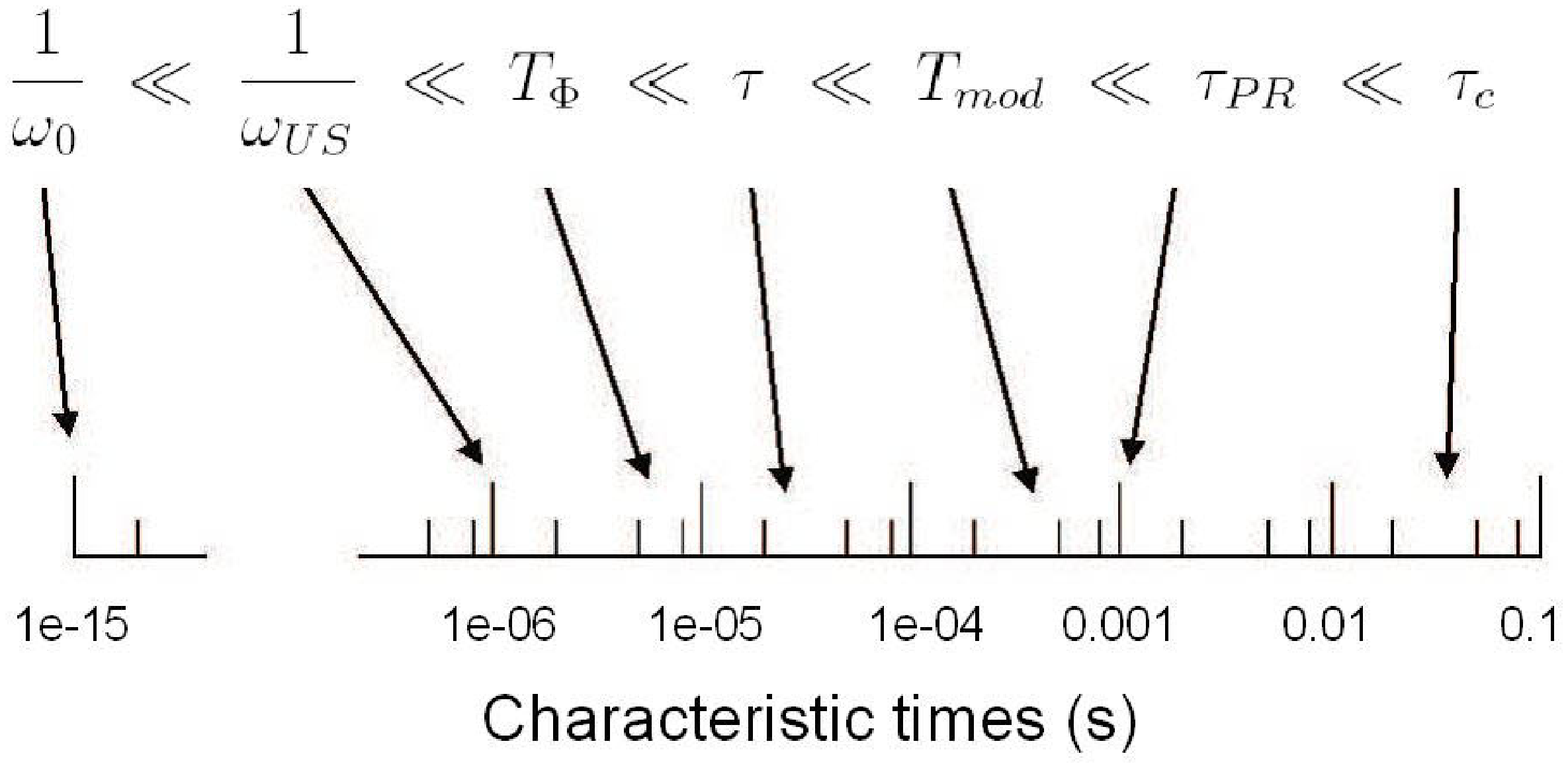}
\caption{
    Order of magnitude of the various times.  $ 1/\omega_{0} $: optical period;
    $ 1 /\omega_{US}$: acoustic period;
    $ T_{\Phi}$: acousto-optical correlation time;
    $ \tau $: time averaging characteristic time;
    $ T_{mod} $:   characteristic time of the modulation $ H(t) $;
    $\tau_{PR} $: photorefractive  time;
    $\tau_{c} $: Lock-in integration time.}
\label{Fig1:fig_times}
\end{figure}
%

%
The incoming optical field and the PZT excitation voltage are now:
\begin{equation}
E'_{P}(t)=\textbf{\ensuremath{\Re}}~\left\{ {\underline{E}}_{P}e^{j(\omega_{0}t+\psi_{P}(t)~)}\right\} \label{Eq_30}\end{equation}
\begin{equation}
U'_{PZT}(t)=\Re~\left\{ {\underline{U}}_{PZT}~e^{j(\omega_{US}t-\psi_{US}(t)~)}\right\}
\label{Eq_31}\end{equation}
where $\psi_{P}$ et $\psi_{US}$ are  random phase modulations applied to the optical
incoming beam $E'_P$ and to the PZT that generates the US beam. Since we consider the
effect  of a random phase modulation,  fields are  noted $E'_{P}$, $E'_{S}$ . The random
phases $\psi_{P}(t)$ and $\psi_{US}(t)$ are supposed to be fully correlated as follow:
%
\begin{equation}\label{Eq_psi}
   \psi_{P}\left(t\right)=\psi_{US}\left(t-z_0/c_{US}\right)
\end{equation}
where $z_0/c_{US}=\theta$ is a fixed temporal delay which determines the $z$ selected
zone $z\simeq z_0$.

To simplify the discussion we will consider here, like in \cite{lesaffre2009acousto},
that the US phase $\psi_{US}$ is randomly drawn every $T_{\Phi}$ to be 0 or $\pi$ with
equal probability. The optical phase $\psi_{P}$ follows the same random phase law than
$\psi_{US}$, but the phase is dealyed in  time by $z_0/c_{US}$. The incoming complex
field is:
%
%

\begin{equation}
{\underline{E}'}_{P}(t)={\underline{E}}_{P}e^{j\psi_{P}(t)~}
\label{Eq_36}
\end{equation}
and the US excitation (${\underline{U}}'_{PZT}$), and  US pressure ($
{\underline{P}}'_{US}$)  complex amplitudes are:
%
%
\begin{eqnarray}\label{Eq_34}
{\underline{U}}'_{PZT}(t)&=&~{\underline{U}}_{PZT}~e^{j\psi_{US}(t)}\\
\nonumber
 {\underline{P}}'_{US}(\textbf{r},t)&=&A(\textbf{r})~{\underline{U}}_{PZT}
\;e^{{j\omega_{US}z}t/{c_{US}}}\;e^{~j\psi_{US}(t-{z}/{c_{US}})}
\end{eqnarray}

\subsubsection{The "tagged photons" field.}
%

By making the calculations leading to Eq.\ref{Eq_22} with the random phases $\psi_{P}$
and  $\psi_{US}$, we get the tagged photons complex amplitude
${\underline{E}}'_{S,\omega_{\pm1}}$:
\begin{eqnarray}\label{Eq_37}
&&{\underline{E}}'_{S,\omega_{\pm1}}(t)=a{\underline{E}}_{P}~\sum_{l} \left[j~e^{-j{2\pi s_{l,0}}/{\lambda}}
\vphantom{\frac{2\pi\beta_{l,m}} {\lambda} \sum_{m}}
\right. \\
\nonumber && \left.
\times \sum_{m} \frac{2\pi\beta_{l,m}} {\lambda}
e^{~\pm j\phi_{l,m}}
~e^{~\pm j\psi_{l,m}(t)}~
\right]
\end{eqnarray}
%
where the phase $\psi_{l, m} $, which depends on time $t$, and on location $z_{l, m}$ of
the $m^{th}$ scatterer along the axis $z$, is defined by:
\begin{eqnarray}\label{Eq_37}
\psi{}_{l,m}(t)=\psi_{P}(t)+\psi_{US}\left(t-{z_{l,m}}/{c_{US}}\right).
\end{eqnarray}
Because of the  random phase jumps, which occur  every $T_\phi$, the complex field
${\underline{E}}'_{S,\omega_{\pm1}}(t)$ varies with a  characteristic time $T_\phi$,
while, in absence of random  modulation, the field ${\underline{E}}_{S,\omega_{\pm1}}$
does not depend on time. In the following, we will detect the field
${\underline{E}}'_{S,\omega_{\pm1}}(t)$ by photorefractive  effect on a crystal.

We must notice that all the photorefractive detection
${\underline{E}}'_{S,\omega_{\pm1}}(t)$ processes occur  on times much larger than
$T_\phi$.
\begin{itemize}
    \item {
The writing of  photorefractive signal on the crystal occurs in a time $\tau_{PR} \gg  T_\phi$.
    }
    \item {
To get a modulated signal for the Lock-In amplifier, the phase of the US will be modulated at a frequency
$\omega_{mod}=1/T_{mod}$ with $T_{mod} \gg T_\phi$.
    }
    \item {
The extraction of the modulated signal modulated at $\omega_{ mod } $ will be made via Lock-In with an integration time
$\tau_c \gg T_\phi$.
    }
\end{itemize}
%
%
So one  can  replace in the following the field ${\underline{E}'}_{S,\omega_{\pm1}}(t)$
by its temporal average $\langle {\underline {E} '}_{, \omega_{\pm1}} (t) \rangle_{\tau}
$ over the characteristic time $\tau$ chosen such as (see Fig.\ref{Fig1:fig_times}):
\begin{equation}\label{Eq_tau}
 T_\phi \ll \tau  \ll \tau_{PR},\; T_{mod}, \;\tau_c
\end{equation}
%
Thus we  eliminate the  fast varying components of ${\underline{E}}'_{S,\omega_{\pm1}}(t)$  which anyway will have no
effect on the final signal.
To be complete let's define here the   temporal average operator $\langle \; \rangle_{\tau}$ :
%
%
\begin{equation}\label{Eq_moy_tau}
    \langle .... \rangle_{\tau} \equiv \frac{1}{\tau} \int_{t'=t-\tau/2}^{t'=t+\tau/2}(....) dt'
\end{equation}
%
The temporal average of the tagged photon field over  the characteristic time $\tau$ is then:
\begin{eqnarray}\label{Eq_39}
 \nonumber && \langle{\underline{E}'}_{S,\omega_{\pm1}}(t)\rangle_{\tau}  =
 a{\underline{E}}_{P}~\sum_{l}\left[j~e^{-j{2\pi s_{l,0}}/{\lambda}} \phantom{\sum_{m} \frac{2\pi\beta_{l,m}}
 {\lambda} } \right.\\
 &&~~~~ \left.\times\sum_{m}\left(\frac{2\pi\beta_{l,m}}
 {\lambda}e^{\pm j\phi_{l,m}}\times\langle e^{\pm j\psi_{l,m}(t)}
 \rangle_{\tau}\right)\right]
 \end{eqnarray}
%
As we can see on Eq.\ref{Eq_39}, $\psi_{l,m}$ acts on the  temporal average
$\langle{\underline{E}'}_{S,\omega_{\pm1}}(t)\rangle_{\tau}$  only through  $\langle
e^{\pm j\psi_{l,m}(t)} \rangle_{\tau}$, which depends only on the  location  along $z$ of
the scatterer of indexes $l,m$, i.e. on $z_{l,m}$.
%
%
From Eq.\ref{Eq_psi} and Eq.\ref{Eq_37}, we have
\begin{eqnarray}\label{Eq_psi pour zlm=z0}
  \nonumber  \psi_{l,m}(t) &\simeq &0 \\
\langle e^{~\pm j\psi_{l,m}(t)} \rangle_{\tau} &\simeq& 1
\end{eqnarray}
\begin{eqnarray}\label{Eq_psi 0 PI}
 \nonumber  \psi_{l,m}(t) &\simeq& 0,  \pi ~~~\textrm{randomly}\\
 \langle e^{\pm j\psi_{l,m}(t)} \rangle_{\tau} &\simeq& 0
\end{eqnarray}
for the scatterer $l, m$ located in (Eq.\ref{Eq_psi pour zlm=z0}) and out (Eq.\ref{Eq_psi
0 PI}) the selected zone $z\simeq z_0$  respectively.

\begin{figure}
  \begin{center}
  \includegraphics[width=8cm]{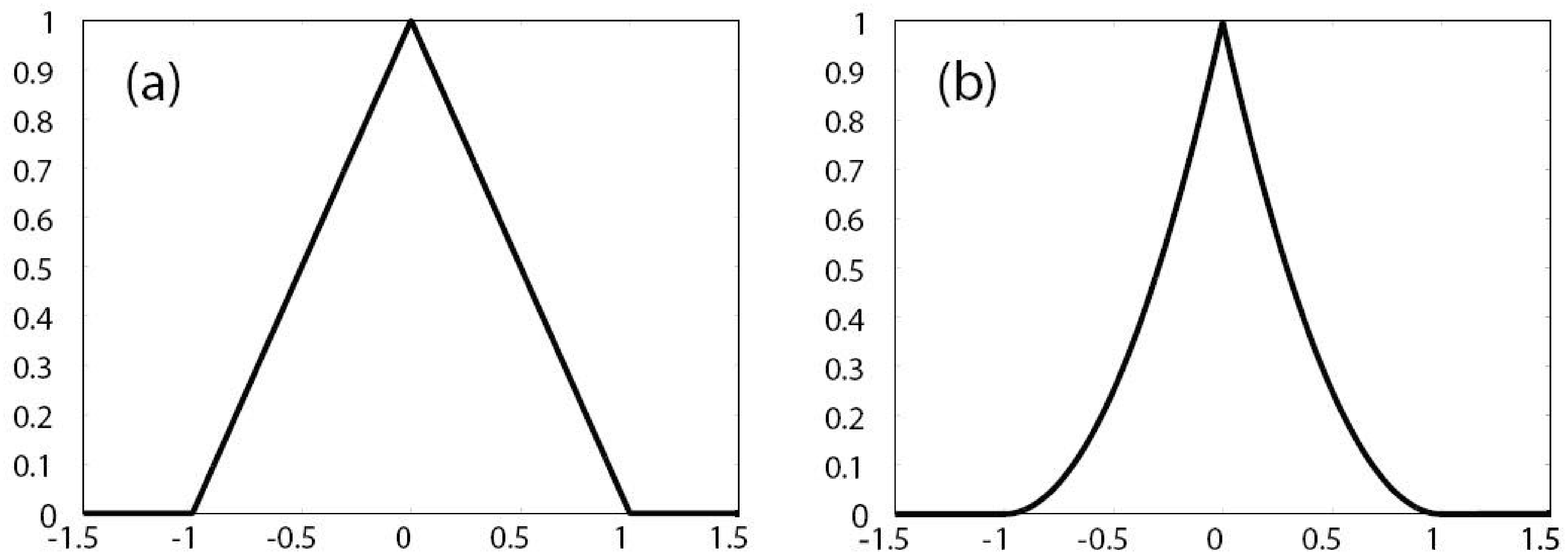}\\
  \caption{Plot of correlation function  $\underline{g}_{1}(\theta)$ (a), and  its square  $|\underline{g}_{1}(\theta)|^2$ (b).
The horizontal axis Units is either $\theta/T_{\Phi}$ (for time correlation), or $(z-z_0)/\Delta z$ with $\Delta z=  c_{US}~T_{\Phi}$
(for $z$ resolution).}\label{Fig_correlation}
 \end{center}
\end{figure}

To characterize this $z$ selection mechanism in a more quantitative way, let us define
the time correlation function:
\begin{eqnarray}
\underline{g}_{1}(\theta) & = & \frac{\langle{\underline{E}}'_{P}(t){\underline{E}}{}_{P}^{\prime*}
(t+\theta)\rangle_{\tau}}{\langle\left|{\underline{E}}'_{P}(t)\right|^{2}\rangle_{\tau}}\label{Eq_32a}\\
 & = & \left\langle e^{j\psi_{P}(t)} e^{-j\psi_{P}(t+\theta)}\right\rangle_{\tau}\nonumber \\
 & = & \left\langle e^{j\psi_{US}(t)} e^{-j\psi_{US}(t+\theta)} \right\rangle_{\tau}\nonumber \\
 & = & \frac{\langle{{\underline{U}}'_{US}(t){\underline{U}}{}_{US}^{\prime*}(t+\theta)}\rangle_{\tau}}
 {\langle{\left|{\underline{U}}'_{US}(t)\right|^{2}}\rangle_{\tau}}\nonumber
\end{eqnarray}
%
%
In the case of $0,\pi$ random phase jumps considered here, $\underline{g}_{1}(\theta)$ is
a triangular function that corresponds to the convolution of two rectangles of width
$T_{Phi}$. The correlation function $\underline{g}_{1}(\theta)$ is plotted on
Fig.\ref{Fig_correlation} (a).

The field $\langle{\underline{E}'}_{S,\omega_{\pm1}}\rangle$ can  be expressed as a
function of $\underline{g}_{1}$:
\begin{eqnarray}\label{Eq_39a}
&& \langle{\underline{E}'}_{S,\omega_{\pm1}}(t)\rangle_{\tau} =
a{\underline{E}}_{P}\sum_{l}\left[j~e^{-j{2\pi s_{l,0}}/ {\lambda}} \vphantom{\sum_{m}} \right.
\\
\nonumber &&~~~~\left. \times\sum_{m}\left[\underline{g}_{1}\left(\frac{z_{l,m}-z_0}{c_{US}}\right)
~\frac{2\pi\beta_{l,m}}{\lambda} e^{\pm j\phi_{l,m}}\right]\right]
\end{eqnarray}
%
Let us note here that the second member  of Eq. \ref{Eq_39a} does not depend on time. It
means that, when we apply the random modulations of phase, the field
${\underline{E}'}_{S,\omega_{\pm1}}(t)$ reaches, after a brief transitory regime, a
stationary regime in which the slow field components do not depend on time  any more.

 Furthermore, the results of the calculations do not depend on $\tau$ as soon as the
condition $T_\phi \ll \tau \ll (\tau_{PR},\; T_{mod}, \;\tau_c)$  of Eq.\ref{Eq_tau} is
fulfilled. So one should write:
%
 $
 \langle{\underline{E}'}_{S,\omega_{\pm1}}(t)\rangle_{\tau}\equiv\langle{\underline{E}'}_{S,\omega_{\pm1}}\rangle$.

\subsection{The photorefractive detection of the tagged photons}
\label{section_pr_dectection}

We will  now  consider the photorefractive detection of the tagged photons in order to
quantify the $z$ selection process for the photorefractive  detected signal (and not just
for the tagged field ${\underline{E}'}_{S,\omega_{\pm1}}$ itself). The calculations we
will make are similar to the ones made by Gross at al. \cite{gross2005theoretical},  but
in a slightly different context.

\subsubsection{The  detection principle}
%
%

%

\begin{figure}
\begin{center}
  \includegraphics[width=8 cm]{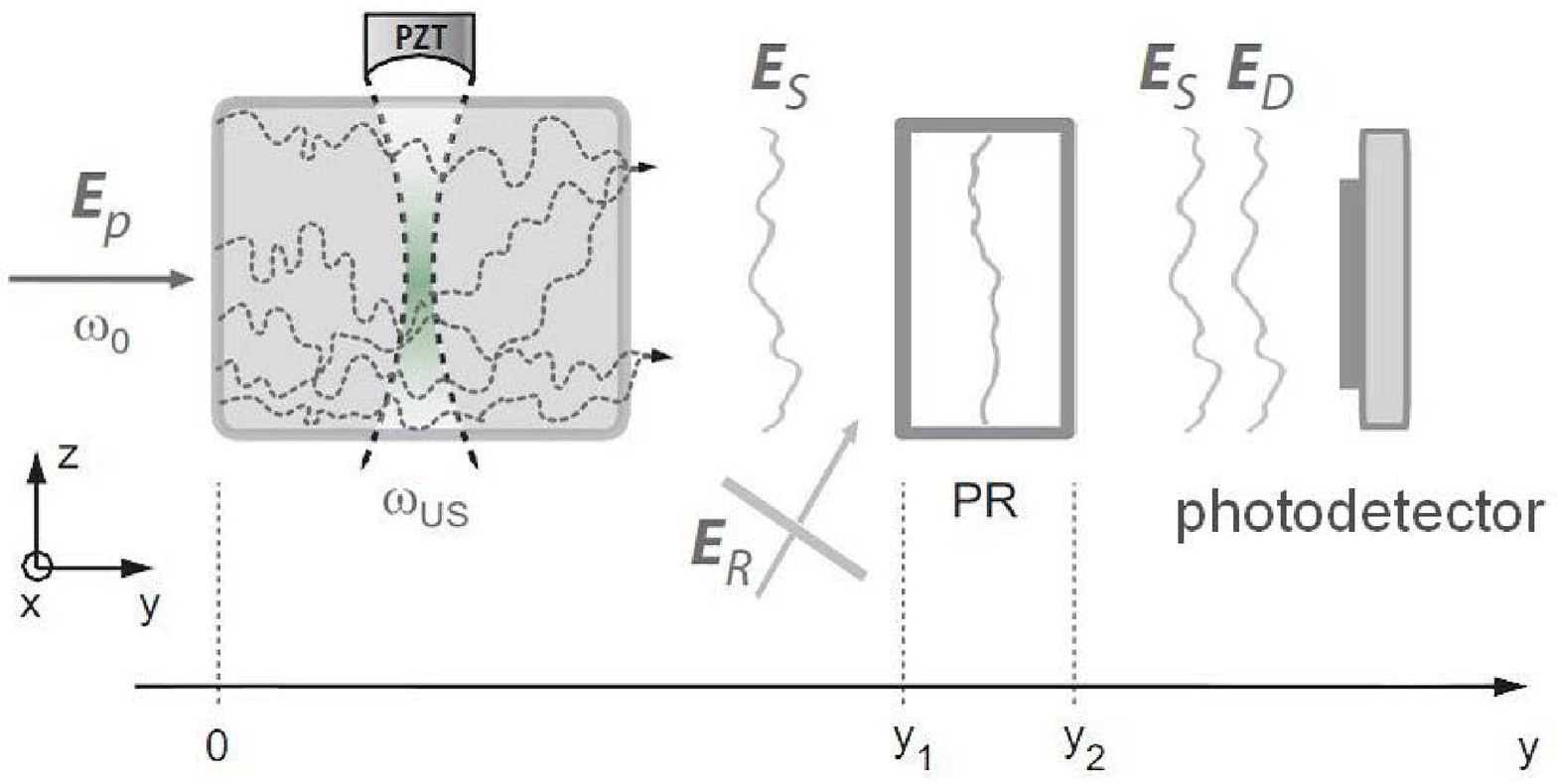}
  \caption{Principle of the photorefractive detection.}\label{fig_PR_detection}
\end{center}
\end{figure}

The principle of the photorefractive detection is illustrated on
Fig.\ref{fig_PR_detection}. The signal $E_{S} $, the wave front of which  is distorted, is
collected in a photorefractive crystal. A reference beam $E_{R}$, considered as plane
wave, which is also called pump beam, interferes with it within  the crystal. By
photorefractive effect, the interferogram grooves a hologram corresponding to a weak
modulation $\delta n(\textbf{r})$ of the local refractive index within the volume of the
crystal. 
This effect having a finite response time $\tau_{PR}$, only the static component of the
interferogram contributes to the recording of the hologram.

To simplify the analysis, we will consider the detection of tagged photons of the $+1$
sideband at $\omega_{1}=\omega_{0}+\omega_{US}$. So we will  shift  the beam reference
frequency by $\omega_{US}=\omega_{1}-\omega_{0}$ in order to perform the photorefractive
detection at frequency $\omega_1$. Let us introduce  the complex amplitude
${\underline{E}}_{R,\omega_{1}}$ of the reference field.
\begin{equation}
E_{R}(t)=\Re~\left[{\underline{E}}_{R,\omega_{1}} e^{j\omega_{1}t}\right]\label{Eq_41}\end{equation}
%
%
The photorefractive effect selects, in the signal field
$E'_{S}=E'_{S,\omega_{0}}+E'_{S,\omega_{1}}+E'_{S,\omega_{-1}}$,  the field component
$E'_{S,\omega_{1}}$. The reference beam is then diffracted by the holographic grating
grooved within the crystal yielding the field $E'_{D}$, whose wavefront is the same for
$E'_{S,\omega_{1}}$. At the exit of the crystal, one gets then both the transmitted
signal beam $E_{S}'$,  and the beam diffracted by the crystal, which will be noted
$E'_{D}$.

\subsubsection{The reference field  diffracted by the crystal $E'_{D}$}


Let us introduce the complex amplitude of the diffracted field defined by
\begin{equation}\label{Eq_E_D__}
E'_{D}(t)=\Re~\left[{\underline{E}}'_{D,\omega_{1}}(t) e^{j\omega_{1}t}\right]
\end{equation}
Let us call  $y=y_{1}$ and $y=y_{2}$ the crystal entrance and  exit planes respectively,
and  $t=0$ the origin of time, when no photorefractive hologram is recorded. Let us
consider that the reference field ${\underline{E}}_{R}$ is  constant.
Within the crystal, the signal field ${\underline{E}}'_{S,\omega_{1}}(y_{1}<y<y_{2},\,
t)$ can be written as a function of the entrance field
${\underline{E}}'_{S,\omega_{1}}(y_{1},t)$
\cite{delaye1995transmission,delaye1997detection}
\begin{eqnarray}\label{Eq_43}
&&{\underline{E}}'_{S,\omega_{1}}(y,t)=  e^{{-\alpha(y-y_{1})}/{2}} \\
&&~~~ \nonumber \times \left[~{\underline{E}}'_{S,\omega_{1}}
(y_{1},t)+\int_{0}^{t}dt'{\underline{E}}'_{S,\omega_{1}}(y_{1},t')G(y,t-t')\right]
\end{eqnarray}
where, under conditions of weak recording efficiency  and weak absorption, the transfer
function $G(y,t)$ can be written as \cite{demontmorillon1997nta}
\begin{equation}
G(y,t)=\frac{\gamma(y-y_{1})}{\tau_{PR}} e^{-\frac{t}{\tau_{PR}}}
\label{Eq_44}\end{equation}
%
Here, $\tau_{PR} $ is the photorefractive response time  and $\gamma$ the photorefractive
gain.
%
%
%
Equation \ref{Eq_43} is established for two plane waves, but it can be generalized to
distorted  wavefront by decomposing the wavefront in plane waves. Several approximations
are made to establish this equation:
%
%
(i) the reference beam is a monochromatic wave with constant frequency
    $\omega_{1}$ (i.e. it is not temporarily  modulated), (ii)
it  is not perturbed  by the recording of the hologram although it can be
    attenuated by the crystal, and (iii)
its power is larger than the signal beam one.
%

In the AOCT experiment \cite{lesaffre2009acousto}, the tagged photon signal is modulated
at a frequency $\omega_{mod}$ of  some kHz, and Lock-in detection is performed. So we are
interested in the low-frequency evolution of $\underline{E}'_{S,\omega_{1}}$.  So we can
replace the $t'=0$  lower limit of the integral $\int dt'$ of Eq. \ref{Eq_43} by
$t'=-\infty$, by neglecting the transient components.  By making the transformation
$t'\rightarrow(t-t')$, in the integral $ \int dt' $ of Eq.\ref {Eq_4}, we obtain:
\begin{eqnarray}\label{Eq_45}
&& \underline{E}'_{S,\omega_{1}}(y,t)=e^{{-\alpha(y-y_{1})}/{2}}\\
\nonumber
&& \times  \left[~{\underline{E}}'_{S,\omega_{1}}(y_{1},t)+\int_{0}^{\infty}{dt'}~
{\underline{E}}'_{S,\omega_{1}}(y_{1},t-t')~G(y,t')~\right]
\end{eqnarray}
%
%
%
We can notice that the hologram is written with delayed time $t-t'$ with a delay $t'$
varying from  zero to  some $\tau_{PR}$. Also let us note that the first term of Eq. \ref
{Eq_45} corresponds to the signal wavefront that is transmitted by  the crystal. Let us
call $E'_{T}$ its  field and  $\underline{E}'_{T}$ its complex amplitude (the index $T$
means here transmitted). The second term corresponds to the reference field that is
diffracted by the crystal  we will  note $E'_{D}$  (where the index $D$ means
diffracted), and $\underline{E}'_{D}$ for the complex amplitude. We can write:
\begin{eqnarray}\label{equ_353a}
E'_{S}(y,t)&=&E'_{T}(y,t)+E'_{D}(y,t)\\
\nonumber E'_{T}(y,t)&=&E'_{T,\omega_{-1}}(y,t)+E'_{T,\omega_{0}}(y,t)+E'_{T,\omega_{+1}}(y,t)\\
\nonumber E'_{D}(y,t)&=&E'_{D,\omega_{+1}}(y,t)
\end{eqnarray}
\begin{eqnarray}\label{equ_353d}
\underline{E}'_{S,\omega_{1}}(y,t)&=&\underline{E}'_{T,\omega_{1}}(y,t)+\underline{E}'_{D,\omega_{1}}(y,t)\\
\nonumber \underline{E}'_{T,\omega_{1}}(y,t)&=&~e^{-\alpha(y-y_{1})/2}{\underline{E}'}_{S,\omega_{1}}(y_{1},t)\\
\nonumber
\underline{E}'_{D,\omega_{1}}(y,t)&=&e^{-\alpha(y-y_{1})/2}\\
\nonumber
&&~~~\times\int_{0}^{\infty}{dt'}{\underline{E}'}_{S,\omega_{1}}(y_{1},t-t')G(y,t')
\end{eqnarray}
Note that since the photorefractive effect selects the field components  of frequency
$\omega_1$, the diffracted field $ E'_{D}$ exhibit  in Eq.\ref{equ_353a} a single
frequency component $E'_{D,\omega_{+1}}$.

\subsection{The acousto optic signal detected by a large area photodiode}
\label{sect:pr_tagged_untagged}

%
%
We consider that the signal is detected by a photodiode of large area located near the
cristal exit plane  $y=y_{2}$. The photodiode signal  $S_{PD}$ is equal to the integral
of  $|\underline{E}'_{S}|^{2}$ over its area.  We get from Eq.\ref{equ_353a}.
\begin{eqnarray}\label{eq:602_}
&&S_{PD}(t)=c.c. +\\
\nonumber && ~~\int dx\int dz~~\left|{\underline{E}'_{S}}(x,y_{2},z,t)\right|^{2}
\end{eqnarray}
\begin{eqnarray}\label{eq:602}
&&S_{PD}(t) = c.c. +\\
 \nonumber &&~~  \int dx\int dz~\left\{~\left|{\underline{E}'_{T}}(x,y_{2},z,t)\right|^{2}  \right.\\
 \nonumber &&~~ \left. \qquad + ~\left|{\underline{E}'_{D}}(x,y_{2},z,t)\right|^{2} \right.\\
\nonumber && ~~\left.   \qquad 
+~\left(~\underline{E}'_{T}(x,y_{2},z,t)~{\underline{E}{}_{D}^{\prime*}}(x,y_{2},z,t)  \right) \right\}
  \end{eqnarray}
%
%
%
%
Because the acousto-optical interaction does not modify the total number of photons, i.e.
 tagged + untagged photons,   the term $|\underline{E}'_{T}|^{2}$ in Eq.\ref{eq:602}
does not depend on the acoustic modulation. Furthermore, because the
gain is supposed to be low, i.e. $\eta_{PR}=\gamma(y_{2}-y_{1})  < 1$, the term
$|\underline{E}'_{D}|^{2}$  can be neglected   in front of the crossed term
$\underline{E}'_{T}~{\underline{E}'_{D}}^{*}$.  Therefore we can only consider the crossed
term, which is the product of the diffracted field  $E_{D}^{\prime}$, which builts up with
the  characteristic time $\tau_{PR}$,  and the transmitted  field $E'_{T}$, which can
vary quickly.  The photodiode modulated signal $S'_{PD}$ is thus:
\begin{eqnarray}
&& S'_{PD}(t)=c.c. + \\
\nonumber && \int dx\int
dz~\underline{E}'_{T,\omega_{1}}(x,y_{2},z,t){\underline{E}{}_{D,\omega_{1}}^{\prime*}}(x,y_{2},z,t)
\end{eqnarray}
From  Eq.\ref{equ_353d}, we get
\begin{eqnarray}\label{equ_603}
&&S'_{PD}(t) = c.c. +\\
\nonumber && e^{-\alpha(y_{2}-y_{1})}\;\int dx\int dz\;{\underline{E}}'_{S,\omega_{1}}(x,y_{1},z,t)\\
 \nonumber && \quad\times\int_{0}^{\infty}{dt'}~{\underline{E}_{S,\omega_{1}}^{\prime*}}
 (x,y_{1},z,t-t')~G^{*}(y_{2},t')
 \end{eqnarray}
%
%
%
To keep a certain universality, we write $G^{*}$ although $G$ is supposed to be real. We
can then develop ${\underline{E}}'_{S,\omega_{1}}$ by summing up all the paths (index $l$)
and scattering events (index $m$) contributions by using Eq.\ref{Eq_39a}. Averaging over
a time $\tau$, we get:
\begin{eqnarray}\label{equ_604}
\nonumber && \langle S'_{PD}(t)\rangle_{\tau}  =  c.c. +  e^{-\alpha(y_{2}-y_{1})}|a\underline{E}_{p}|^{2}\;\int dx\int dz\;\\
 && \left[\sum_{l}j e^{-j{2\pi  s_{l,0}}/{\lambda}}\times\sum_{m}\frac{2\pi\beta_{l,m}}{\lambda}~\underline{g}_{1}
 \left(\frac{z_{0}-z_{l,m}}{v_{US}}\right) e^{j\phi_{l,m}}\right]\nonumber \\
 && \times\left[\int_{0}^{\infty}dt'~~G^{\ast}(y_{2},t')\times\left(\sum_{l'}-j~
 e^{{2j\pi s_{l',0}}/{\lambda}}\right.\right.\nonumber \\
&&  \quad\quad\left.\left.\times\sum_{m'}\frac{2\pi\beta_{l',m'}}{\lambda}\;\underline{g}_{1}^{*}
 \left(\frac{z_{0}-z_{l',m'}}{v_{US}}\right)   ~e^{-j\phi_{l',m'}}\right)\right]\nonumber\\
 \end{eqnarray}
%
%
%
%
The equation \ref{equ_604} illustrates the complexity of the calculation of the  signal.
It involves a double summation over  the optical paths ( i.e. $\sum _ {l,l'} $), a
double summation over the scattering  events  (i.e. $\sum _{m, m '}$), a  spatial
integral over the photodiode area (i.e. $\int\int dx~dz$), and a temporal integral over
the delay $t'$ (i.e. $\int dt' $).
%
%

To simplify this equation, let us consider first the integral over the photodiode area
$\int \int dx \: dz$. Every point ($x, z$) of the photodiode selects paths  $l$ and
$l'$, which finishes  in  ($x, z$). For the corresponding paths, the phase factor
$e^{-j{2\pi\; s_{l,0}}/{\lambda}}$  is totally random from a route to the next one. So we  can
limit the summation over $l$ and $l'$ to  the terms $l=l'$. The equation \ref{equ_604}
becomes then:
\begin{eqnarray}\label{equ_604a}
\nonumber  \langle S'_{PD}(t)\rangle_{\tau} =  c.c.+e^{-\alpha(y_{2}-y_{1})}\left|\frac{2\pi
a\underline{E}_{p}}{\lambda}\right|^{2}\;\int dx\int dz \\
~~\times \sum_{l}\sum_{m}
  \left[\beta_{l,m}\;\underline{g}_{1}\left(\frac{z_{0}-z_{l,m}}{c_{US}}\right)\right.\int_{0}^{\infty}dt'G^{\ast}(y_{2},t')
  \nonumber \\
  ~~\left.\times \sum_{m'}\beta_{l,m'}\;\underline{g}_{1}^{*}\left(\frac{z_{0}-z_{l,m'}}{c_{US}}\right)
 e^{j(\phi_{l,m}-\phi_{l,m'})}\right]
 \end{eqnarray}
%
%
%
%
To simplify  this equation further, it is necessary to study the mutual coherence of the
phases $\phi_{l, m}$ and $\phi_{l, m'}$ that corresponds to two different scattering
events $m$ and $m'$ of the same path $l$. According to the position of the scatterers,
and according to the acousto-optical modulation mechanism, these phases are correlated or
not. Nevertheless,  when the two events ($m$ and $m'$) occur in two $z$ coordinates
$z_{l, m}$ and $z_{l, m'}$ separated by more than an acoustic  wavelength $\lambda_{US}$
(i.e. $ |z_{l, m'}-z _{l, m}| >  \lambda_{US} $), the  phases $\phi_{l, m}$ and $\phi_{l,
m'}$ are weakly correlated. We can then write:

\begin{eqnarray}\label{eq:605}
 &&\langle S'_{PD}(t)\rangle_{\tau} =  c.c.+\\
\nonumber && e^{-\alpha(y_{2}-y_{1})}\left|\frac{2\pi a
\underline{E}_{p}}{\lambda}\right|^{2} \intop_{0}^{\infty}dt'   G^{\ast}(y_{2},t') \int dx\int dz \\
&& \nonumber \sum_{l}\sum_{m}\sum_{m'~\textrm{with} ~\mid z_{l,m}-z_{l,m^{\prime}}\mid<\lambda_{US}}
\left[\beta_{l,m}\beta_{l,m'}  \vphantom{
\underline{g}_{1}^{*}\left(\frac{z_{0}-z_{l,m'}}{c_{US}}\right)}
\right.\\
 \nonumber
 &&  \left. e^{j\left(\phi_{l,m}-\phi_{l,m'}\right)}\underline{g}_{1}\left(\frac{z_{0}-z_{l,m}}{c_{US}}\right)\:
 \underline{g}_{1}^{*}\left(\frac{z_{0}-z_{l,m'}}{c_{US}}\right)\right]
  \end{eqnarray}
%
%
%
Since the magnitude of the acoustic pressure vary weakly over  $\lambda_{US} $, we have
$\beta_{l, m} \simeq \beta_{l, m'}$ for $z_{l, m}-z_{l, m'} < \lambda_{US}$. Moreover,
the random modulation of phases is chosen in such a way that the characteristic length
$T_{\phi}~c_{US} $ is larger than the acoustic wavelength $\lambda_{US} $ (i.e. $T_{\phi}
~c_{US} > \lambda _ {US} $. This implies that $\underline {g}_{1} \left ( (z_{0}-z_{l,
m}) /{c_{US}} \right) \simeq \underline {g}_1 \left ( (z_{0}-z_{l, m '})/ {c_{US}}
\right) $ for $z_{l, m}-z_{l, m'}  < \lambda_{US}$. Therefore we  obtain:
%
\begin{eqnarray}\label{eq:606b1}
&& \langle S'_{PD}(t)\rangle_{\tau}  =  c.c.\\
&& \nonumber +e^{-\alpha(y_{2}-y_{1})}\left|\frac{2\pi a\underline{E}_{p}}{\lambda}\right|^{2}\intop_{0}^{\infty}dt'   G^{\ast}(y_{2},t')  \int dx\int dz\\
 &&\sum_{l}\sum_{m}\beta_{l,m}^{2}\; \left| \underline{g}_{1}\left(\frac{z_{0}-z_{l,m}}{c_{US}}\right)\right|^{2}\nonumber \\
 && \quad \quad  \times\sum_{m'~\textrm{with} ~\mid z_{l,m}-z_{l,m^{\prime}}\mid<\lambda_{US}}e^{j\left(\phi_{l,m}-\phi_{l,m'}\right)} \nonumber
\end{eqnarray}
%
%
%
Here, the term $\left| \underline{g}_{1} \left ({z_{0}-z_{l, m}} /{c_{US}} \right) \right
|^2$ selects the zone of imaging.

The $z$ resolution one can expect is roughly equal to the half-width of
$\underline{g}_{1}(z)$, i.e. to $0.5\times  c_{US} T_{\Phi}$ (see
Fig.\ref{Fig_correlation}). The expected resolution is thus about 7.5 mm for
$T_{\Phi}=10$ $\mu$s (20 US periods at $\omega_{US}=2$ MHz), and 1.1 mm for
$T_{\Phi}=1.5$ $\mu$s (3 US periods at $\omega_{US}=2$ MHz). The AOCT published
experimental results \cite{lesaffre2009acousto} correspond to $T_{\Phi}\simeq 2$ $\mu$s.
To improve the $z$ resolution, one must thus decrease $T_{\Phi}$. The acousto optic
signal decreases accordingly, since it is proportional to $T_{\Phi}$:  the scattering
events that contribute to the signal must be within the $\underline{g}_{1}(z)$ selected
region.

\subsection{The Lock-in detection of the acousto optical signal} \label{PR_sig_with_time_US}


\subsubsection{The modulation of the signal at $\omega_{mod}$}
%

Note that  $\langle S_{PD} ( t ) \rangle_{\tau} $ given by Eq.\ref{eq:606b1} is invariant
with time. So the  tagged photons photorefractive signal $S_{PD}$ is a CW
 component, which adds to the total flow of transmitted light. To detect
$S_{PD}$ more efficiently with a Lock-in, AOCT  adds an extra modulation of the
ultrasonic wave. Like in the AOCT experiment \cite{lesaffre2009acousto},  we will
consider here an asymmetric $0$ to $\pi$ phase modulation $H_{U.S.}(t)$ at frequency  $
\omega_{mod} = 2 \pi / T_{mod} \sim $ 3 kHz, with duty cycle $ 0 <r <1 $:
\begin{eqnarray}
H_{US}(t) & = & +1\quad\quad\textrm{pour}\quad\quad0\leq t/T_{mod}\leq r\label{equ_702}\\
H_{US}(t) & = & -1\quad\quad\textrm{pour}\quad\quad r<t/T_{mod}\leq1\nonumber \end{eqnarray}
%
%
We consider that the modulation period $ T_{mod} $ is very large compared to the
correlation time $ T_{\phi} $, but smaller than the photorefractive time $ \tau_{PR}$,
i.e.
%
$({2\pi}/{\omega_{US}})<T_{\phi}\ll T_{mod}<\tau_{PR}$.
%
In practice, we typically use $ T_{mod} \sim 100 \mu$s  (see Fig.\ref{Fig1:fig_times}).
The modulation is applied according
\begin{equation}\label{equ_701}
\underline{U}'_{PZT}(t)\rightarrow\underline{U}''_{PZT}(t)=~H_{US}(t)~\underline{U}'_{PZT}(t)
\end{equation}
The US signal is denoted $ \underline{U}''_{PZT} (t) $, and the fields
are denoted $ E''_{P} $, $E''_{S} $  and so on. 
The complex amplitude of the tagged  photon field is now
\begin{equation}\label{equ_704}
\underline{E}'_{S,\omega_{1}}(t)\rightarrow\underline{E}''_{S,\omega_{1}}(t)=~H(t)~\underline{E}'_{S,\omega_{1}}(t)
\end{equation}
where $H(t)=H_{US}(t-z_{0}/c_{US})$.
%
%
%
Similarly with Eq.\ref {equ_353d},   the diffracted complex amplitude  becomes
\begin{eqnarray}\label{equ_705}
\nonumber  \underline{E}''_{D,\omega_{1}}(y,t)&=&e^{-\alpha(y-y_{1})/2} \int_{0}^{\infty}{dt'} \left[H(t-t')\right. \\
 && \left. \times {\underline{E}'}_{S,\omega_{1}}(y_{1},t-t')~G(y,t') \right]
\end{eqnarray}

\subsubsection{The modulated acousto-optical signal}

%
The signal from the photodiode given by equation \ref {equ_603} becomes
\begin{eqnarray}\label{equ_706}
&& S''_{PD}(t) =  c.c.\\
&& \nonumber  \vphantom{\int dx} +~e^{-\alpha(y_{2}-y_{1})}\;\int dx\int dz \left[H(t)\vphantom{\int dx} {\underline{E}}'_{S,\omega_{1}}(x,y_{2},z,t)\right.\\
 && \left.
 \int_{0}^{\infty}{dt'}~H(t-t')\;{\underline{E}^{\prime*}{}_{S,\omega_{1}}}(x,y_{2},z,t-t')~G^{*}(y_{2},t')
 \right]\nonumber
 \end{eqnarray}
%
%
%
By making the calculation  leading to Eq. \ref{eq:606b1} from to Eq. \ref{equ_603} with the
additional modulation $H_{US}$, we get.
\begin{eqnarray}\label{eq:606b2}
&&\langle S''_{PD}(t)\rangle_{\tau} =  c.c.+ \\
\nonumber && e^{-\alpha(y_{2}-y_{1})}\left|\frac{2\pi a\underline{E}_{p}}{\lambda}\right|^{2}H\left(t\right) \intop_{0}^{\infty}dt'H\left(t-t^{\prime}\right)G^{\ast}(y_{2},t')\nonumber \\
 &  & \times\int dx\int dz\sum_{l}\sum_{m}\left[\beta_{l,m}^{2}\;\left|\underline{g}_{1}\left(\frac{z_{0}-z_{l,m}}{c_{US}}\right)\right|^{2} \right.\nonumber \\
 &  & \left.\times\sum_{m'/\mid z_{l,m}-z_{l,m^{\prime}}\mid<\lambda_{US}}e^{j\left(\phi_{l,m}-\phi_{l,m'}\right)}\right]\nonumber \end{eqnarray}
%
%
%
Since we have  consider $ T_{mod} < \tau_{PR} $, the integration over $ t'$ can be
simplified, and we obtain from Eq.\ref{Eq_44}.
\begin{eqnarray}\label{equ_709}
\int_{0}^{\infty}dt'H(t-t')\; G^{\ast}(y_{2},t')= ~~~~~~~~~~~~~~~~~~~~~~~~~\\
\nonumber ~~~~\left[\frac{1}{T_{PR}}\int_{0}^{T_{PR}}dt'H(t-t')\right] \times\left[\int_{0}^{\infty}dt'G^{\ast}(y_{2},t')\right]\nonumber \\
 \nonumber ~~~~~~~~~~~~~~~~~~~~~~~~~~~~~~~~~~~~~= (1-2r)\;\gamma(y_{2}-y_{1})
 \end{eqnarray}
%
%
%
This means that for a modulation $ H(t) $ faster than  $\tau_{PR} $, the photorefractive
recorded hologram is proportional to the average $ \langle H (t ) \rangle_{\tau_{PR}}$.
The asymmetric nature of the modulation $ H \left (t \right) $ yields non-zero
photorefractive grating. The modulated component of the   signal on a large area
photodiode thus becomes.
\begin{eqnarray}\label{eq:606b3}
& & \langle S''_{PD}(t)\rangle_{\tau} \simeq c.c + \\
&  & e^{-\alpha(y_{2}-y_{1})}\left|\frac{2\pi a\underline{E}_{p}}{\lambda}\right|^{2}H\left(t\right) (1-2r)\;\gamma(y_{2}-y_{1})\nonumber \\
 &  & \times\int dx\int dz\sum_{l}\sum_{m}\beta_{l,m}^{2}\left| \underline{g}_{1}\left(\frac{z_{0}-z_{l,m}}{c_{US}}\right)\right|^{2}\nonumber \\
 &  & \times\sum_{m'/\mid z_{l,m}-z_{l,m^{\prime}}\mid<\lambda_{US}}e^{j\left(\phi_{l,m}-\phi_{l,m'}\right)}\nonumber
 \end{eqnarray}
By this way, the photodiode signal is modulated following  $H(t)$.
%


%
Equation \ref{eq:606b3}, which does not depend on $\tau$ can be slightly simplified as
following:
\begin{eqnarray}\label{eq:606b4}
\nonumber &&\langle S''_{PD}(t)\rangle  = (1-2r)H(t)\!\left|\underline{E}_{p}\right|^{2}\;
\gamma(y_{2}-y_{1})e^{-\alpha(y_{2}-y_{1})}\\
 && ~~~~\int dx\int dz\sum_{l}\sum_{m}\left[\beta_{l,m}^{2}\left|\underline{g}_{1}\left(\frac{z_{0}-z_{l,m}}{c_{US}}\right)\right|^{2} \vphantom{\sum_{m'/\mid z_{l,m}-z_{l,m^{\prime}}\mid<\lambda_{US}}e^X}\right.\nonumber \\
 && ~~~~\left. \sum_{m' ~\textrm{with} ~\mid z_{l,m}-z_{l,m^{\prime}}\mid<\lambda_{US}}\left(e^{j\left(\phi_{l,m}-\phi_{l,m'}    \right)}+c.c. \right) \right]\nonumber\\
 \end{eqnarray}
%
%
%
%

\section{Conclusion}

The mains results of the paper are  Eq.\ref{Eq_39a}  and Eq.\ref{eq:606b4}. Equation
\ref{Eq_39a} shows that the tagged photon field $\langle E'_{S,\omega_{\pm 1}}\rangle$,
which
 is
 \begin{itemize}

 \item {proportional to the amplitude of the optical field  $ E_{P} $ injected in the
     scattering medium,}

 \item {proportional to the acoustic power delivered by the PZT via the term $\beta_
     {l, m}^{2}$}

    \item { and proportional to the correlation function $\underline {g}_{1}
        ({z_{0}-z_{l, m}} /{c_ {US}})$. }
\end{itemize}
The random phase modulation  creates along the acoustic column a zone of coherence
located near $z \simeq z_0 = c_{US} \theta$. The tagged photon signal from this zone adds
up coherently, and can be further detected. For the tagged photon field,  the random
phase jump $z$ selection is quantified  by the factor $\underline {g}_{1} ({z_{0}-z_{l,
m}}/c_{US})$, which is the correlation product of a rectangle of width $T_{\phi}$. This
correlation product, which  has triangular shape, is  plotted on
Fig.\ref{Fig_correlation}(a) as a function of time, in $T_{\Phi}$ Units, or as a function
of the scatterer $z$ relative coordinate (i.e. ${z_{0}-z_{l, m}}$), in $\Delta z= c_{US}
T_{\Phi}$ Units.

%
%
%
%
%
%

On the other hand, Eq.\ref{eq:606b4} shows that the  acousto-optical modulated signal  on
the large area photodiode surface $\langle S_{PD}(t)\rangle_{\tau} $ is
 \begin{itemize}

 \item {proportional to the optical intensity  $ |E_{P}|^2 $ injected in the
     scattering medium,}

 \item {proportional to the surface of the photodiode, i.e. $\int dx\int dz$,}

 \item {proportional to the acoustic power delivered by the PZT via the term $\beta_
     {l, m}^{2}$}

 \item {proportional to  $ (1-2r) H(t)$ where $ H(t) $ is the additional time
     modulation, whose  duty cycle is $r$. Because of this well controlled time
     modulation, the  photodiode signal can be Lock-in detected at the frequency $
     \omega_ {mod} $ with an integration time $ \tau_{c}> T_{mod} = 2 \pi /
     \omega_{mod} $.}

    \item { and proportional to the square of the correlation function
        $\underline{g}_{1}$, i.e. to $\left|\underline {g}_{1} ({z_{0}-z_{l, m}} /{c_
        {US}} )\right|^{2} $. }
\end{itemize}
For the photodiode signal,  the random phase jump $z$ selection is quantified  by the
factor $\left|\underline {g}_{1} ({z_{0}-z_{l, m}} /{c_
        {US}} )\right|^{2} $, which is the square of the correlation product
        $\underline{g}_{1}$. This $|\underline{g}_{1}|^2$ factor is  plotted on Fig.\ref{Fig_correlation}(b).
One must note also that the summation over $m'$ of the phases factors
$e^{j\left(\phi_{l,m}-\phi_{l,m'}\right)}$, which can be limited to  $| z_{l,m}-z_{l,m'}|
<\lambda_{US}$,  describes  here the effect of  partial coherence of the successive
scattering events within a given travel path $l$. The corresponding proportionality
factor does not depend on the US modulation, and does not provide any $z$ selection.

In this paper, we have described the AOCT effect theoretically.   We show that the tagged
photons remain coherent if they are generated within a selected $z$ region of the sample,
and we have quantified this $z$ selection effect for both the tagged photon field $
E'_{S,\omega_{\pm 1}}$, and the photorefractive photodiode signal  $S''_{PD}(t)$.  These
theoretical results will be compared with experiment in another publication.

\bibliographystyle{osajnl}

\end{document}